# An Empirical Study of Safetensors' Usage Trends and Developers' Perceptions

Beatrice Casey, Kaia Damian, Andrew Cotaj, and Joanna C. S. Santos
*University of Notre Dame*
Notre Dame, IN, USA
{bcasey6, kdamian, acotaj, joannacss}@nd.edu

***Abstract*—Developers are sharing pre-trained Machine Learning (ML) models through a variety of model sharing platforms, such as Hugging Face, in an effort to make ML development more collaborative. To share the models, they must first be serialized. While there are many methods of serialization in Python, most of them are unsafe. To tame this insecurity, Hugging Face released safetensors as a way to mitigate the threats posed by unsafe serialization formats. In this context, this paper investigates developer's shifts towards using safetensors on Hugging Face in an effort to understand security practices in the ML development community, as well as how developers react to new methods of serialization. Our results find that more developers are adopting safetensors, and many safetensor adoptions were made by automated conversions of existing models by Hugging Face's conversion tool. We also found, however, that a majority of developers ignore the conversion tool's pull requests, and that while many developers are facing issues with using safetensors, they are eager to learn about and adapt the format.**

## I. Introduction

Pre-trained models have recently risen in popularity, as platforms such as Hugging Face [1] aim to democratize and and further advance Artificial Intelligence (AI) and Machine Learning (ML) development through collaboration among developers. Hugging Face acts as a hub, where developers can upload and share their models. Developers can then download these pre-trained models, and either fine tune them on downstream tasks [2]–[6], or directly integrate them into their systems. Pre-trained models allow for model development to become more efficient and less expensive, allowing developers to further their innovation [7].

**Model serialization** is crucial for using pre-trained models, as it allows developers to save and share a model, along with all of its parameters, on these model hubs [8]. While serialization is crucial for increasing the efficiency of software development, it also has introduced new vulnerabilities, such as AI supply chain attacks [9], [10]. In such attacks, an adversary replaces a *benign* model with a *malicious* one that contains a payload to be executed when the model is loaded [11]. These supply chain attacks are a result of using *unsafe* serialization formats. While models can be serialized using a multitude of serialization formats (*e.g.*, Pickle), the majority of these formats allow arbitrary code execution during model deserialization [12]. To tame the inherent unsafety of existing serialization formats, Hugging Face introduced **safetensors** [13] as a novel, safe method of serialization aimed at circumventing the threat of object injection vulnerabilities in ML models. This format was specifically designed to not only prevent vulnerabilities but also to improve model loading times [14]. To facilitate the adoption of this new format, Hugging Face also released a tool [15] that automatically converts models saved using PyTorch to the safetensors format.

While a recent study [16] provides preliminary evidence that developers have begun shifting towards using safetensors, it is unclear how fast/slow this shift happened, how developers reacted to the release of this format and what are the technical challenges faced by developers that hinder safetensors' adoption. Thus, we aim to understand the shift to safetensors in an effort to uncover evolving security practices in the machine learning community and the willingness to adopt safer alternatives to traditional formats, such as pickle and PyTorch's default tensor storage. As safetensors mitigate security risks associated with traditional model formats, which can be prone to malicious payload injection, understanding adoption rates helps assess whether developers prioritize security in model storage and informs future recommendations on safe model serialization practices.

In light of this, in this paper, we investigate developers' shifts towards using safetensors on Hugging Face, as well as their sentiments towards the format. We perform a large-scale empirical study of open source AI/ML models on Hugging Face. To conduct this study, we collected and analyzed the commit history from Hugging Face model repositories, identifying the serialization formats used to save models. We then extract all pull requests (PRs) that were made as a result of using Hugging Face's conversion tool. Additionally, we searched developer forums (*e.g.*, Stack Overflow, GitHub) for discussions related to SafeTensors. We conducted open coding on these discussions to identify key concepts and themes, creating a taxonomy of developers' perceptions on safetensors.

In our study, we found a general interest in the format with safetensors gaining traction, due to its security and efficiency. We also found that developers face a variety of issues when adopting safetensors, such errors when loading the model and issues with model performance after converting to safetensors.

This paper makes two major contributions:

- ***A Study of Serialization Formats Usage Evolution***: We empirically study the frequency of use for different serialization formats across Hugging Face to determine the shift towards using safetensors, a safer serialization format.

- ***Analysis of Developer Sentiments***: We collect and analyze developers' sentiments about the safetensors formats through a systematic analysis of developers' discussions. This resulted in a taxonomy of developers' perceptions and concerns related to using safetensors.

The rest of this paper is organized as follows: Section II provides a background on the key concepts necessary to understand our work. Section III explains the methodology of this work. Section IV shares the results from our study. Section V provides a disccusion of our findings. Section VI describes related work, Section VII shares the threats to validity, and Section VIII concludes the work. This study's replication package is available in an anonymized GitHub [17].

## II. Background

This section describes core concepts such that the work can be understood by a broader audience.

### A. Model Serialization and Deserialization

**Model serialization** is the process of converting a model to *abstract representations*, such as binary or JSON formats [18]–[21]. Conversely, model deserializaiton is the process of restoring a model from its abstract representation [22]. Model serialization is a key step in the development of ML-based software systems [16]. It allows model sharing and reuse, reducing the need to retrain or rebuild models every time they need to be used [8]. This way, one can save (*serialize*) a model to save its architecture, weights, hyperparameters, *etc* and load (*deserialize*) it when the model is needed for inference.

### B. Model Serialization and Security

Although many programming languages can be used to develop machine learning models, Python remains the leader in ML development [23]. There are many ways which a model can be serialized in Python, depending on the needs of the developer. However, certain methods of serialization (namely `torch.save`, `pickle`, `joblib`, `dill`, `ONNX`, `numpy`, `H5/HDF5`, and `torch.jit.save`) are vulnerable to **object injection vulnerabilities**, where an attacker is able to inject arbitrary code into the bytes of a file that execute upon deserialization [12]. Prior work has shown that this problem [16] is rampant on Hugging Face.

train.py
```python
import pickle
import torch
import torch.nn as nn
import torch.optim as optim
from utils import MyNN

# Train a simple Neural Network
model = MyNN() # model from utils.py
...
# store the model using torch .
torch.save(model, 'my_model.bin')
```

inference.py
```python
import pickle
import torch
...
# Load model saved using torch .
model = torch.load('my_model.bin') .
path = "example.jpg"
image = preprocess_image(path)
with torch.no_grad():
  output = model(image)
  _, p = torch.max(output, 1)[1].item()
  prediction = p.item()
```

Listing 1: Example of Model De/Serialization

Listing 1 demonstrates an example of insecure model deserialization. On the left, a model is trained and then serialized using PyTorch's *toch.save* which produces a pickle-based serialized model file (`my_model.bin`). Then, in `inference.py`, the model is deserialized (*e.g.*, loaded) to be used on the example image. If the model contained a malicious payload (such as in Figure 1), this payload would be executed on the developer's machine during model loading. Figure 1 demonstrates the bytes of the pickle file for a benign and a malicious model. On the left is the benign model, where the model created in Listing 1 is loaded. However, on the right is a malicious model, where an attacker embedded arbitrary code to execute the command 'rm -rf /'. This code was embedded at the front of the file, meaning when the developer loads the model file as in Listing 1, this payload would execute, and the developer would have all the files on their machine removed.

Original Model
```
0: \x80 PROTO      4
2: \x95 FRAME      676
11: \x8c SHORT_BINUNICODE 'utils'
18: \x94 MEMOIZE    (as 0)
19: \x8c SHORT_BINUNICODE 'MyNN'
29: \x94 MEMOIZE    (as 1)
30: \x93 STACK_GLOBAL
31: \x94 MEMOIZE    (as 2)
32: )    EMPTY_TUPLE
33: \x81 NEWOBJ
34: \x94 MEMOIZE    (as 3)
35: }    EMPTY_DICT
36: \x94 MEMOIZE    (as 4)
37: (    MARK
38: \x8c     SHORT_BINUNICODE 'training'
48: \x94     MEMOIZE    (as 5)
49: \x88     NEWTRUE
...
```

Malicious Model
```
0: \x80 PROTO      5
2: \x95 FRAME      758
11: \x8c SHORT_BINUNICODE 'builtins'
21: \x94 MEMOIZE    (as 0)
22: \x8c SHORT_BINUNICODE 'exec'
28: \x94 MEMOIZE    (as 1)
29: \x93 STACK_GLOBAL
30: \x94 MEMOIZE    (as 2)
31: \x8c SHORT_BINUNICODE "rm -rf /"
88: \x94 MEMOIZE    (as 3)
89: \x85 TUPLE1
90: \x94 MEMOIZE    (as 4)
91: R    REDUCE
92: \x94 MEMOIZE    (as 5)
93: \x8c SHORT_BINUNICODE 'utils'
100: \x94 MEMOIZE    (as 6)
101: \x8c SHORT_BINUNICODE 'MyNN'
...
```

Fig. 1. Decompiled Pickle Model (benign and malicious versions)

### C. Safetensors

Hugging Face released in 2022 a new serialization format, named *safetensors*, as a solution to the problem of object injection [13]. This format aims to not only allow safe serialization but also is designed to be optimized for quick loading and reading. This format mitigates deserialization vulnerabilities by only allowing a developer to save numeric information (such as weights and parameters), but *not* allowing callbacks, model architecture, or any other forms of arbitrary code execution. As shown in Figure 2, safetensors' format first contains the size of the header, then the header in a JSON format, with information such as tensor names and data types, *etc*. The rest of the file follows in binary. By only allowing for numeric information to be saved, safetensors effectively avoids the issue of object injection. However, this format greatly limits what information a developer is allowed to save in their model file, which might impact a developer's decision to choose this format.

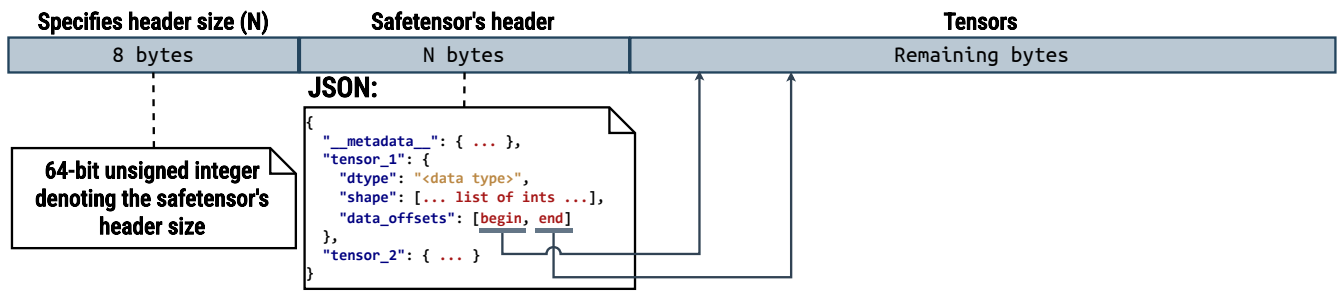

Fig. 2. Safetensor serialization format

### D. Hugging Face's Safetensor Conversion Tool

Hugging Face has an online tool [15] that provides a user-friendly way to convert models serialized using PyTorch to the safetensors format. The conversion process is straightforward: developers specify the URL to their repository containing the serialized model in the PyTorch format, and the tool automatically transforms it into the Safetensors format. Once the conversion is complete, the tool opens a pull request (PR) to the repository where the original model is located, and developers can choose whether to accept the merge or close/ignore the request. This PR contains a new file (*model.safetensors*) that retains all the original model's data while offering enhanced security benefits.

It is important to highlight that anyone can use this converter tool to convert not only models in repositories they own but also for any public repository. This means that a public repository may receive pull requests from this conversion tool that were triggered by external developers.

## III. Methodology

This section explains our research questions (RQs), and the methodology we followed to answer these questions. Figure 3 provides an overview of how we answer each of these RQs.

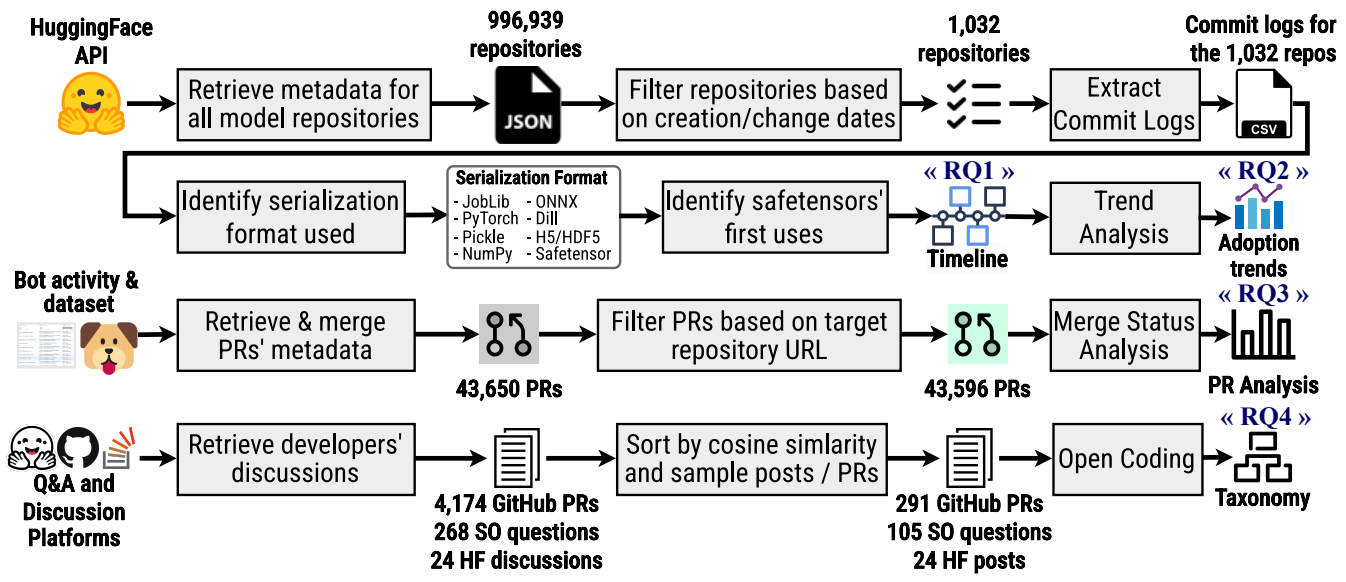

Fig. 3. Study Overview

### A. Research Questions

This paper answers the following research questions:

**RQ1** ***How long it took developers to adopt safetensors in their projects?***

While safetensors was first released in September 2022 [24], we do not currently have an understanding of how fast safetensors was adopted as a serialization format for models. In this question, we aim to understand when developers began using safetensors in their repositories.

**RQ2** ***What is the trend of adoption of safetensors?***

While RQ1 focuses on finding the point in time which safetensors started being used, in RQ2 we examine at what *speed* safetensors have been adopted. In other words, we investigate whether developers gradually migrated to this format or quickly adopted it.

**RQ3** ***How effective is Hugging Face's conversion tool in converting PyTorch models to safetensors?***

In this RQ, we examine Hugging Face's conversion tool effectiveness. Specifically, we scrutinize the PRs created by the tool that were submitted to the original repositories and study their merge rates. This question aims to gather insights into the community's responsiveness and openness to safetensors as a secure and efficient alternative. Additionally, we analyze comments and discussion in these PRs to identify external factors that hinder safetensors adoption, such as developers' trust in the format or any difficulties experienced by developers during the conversion.

**RQ4** ***What are developers perceptions on safetensors?***

This question focused on understanding how developers adapted to safetensors, as well as finding common themes or sentiments in discussions that developers have. We answer this question by performing a qualitative analysis of developers' discussions posted on StackOverflow, GitHub and Hugging Face to understand the way developers' view safetensors.

### B. Model Repository Metadata Retrieval

Since Hugging Face [1] is currently the most popular online model hub [7], as well as the company which created the safetensors format [13], we use it as our hub for model collection and analysis for this study. The Hugging Face Hub hosts three repository types [25]: *model repositories*, *dataset repositories*, and *spaces*, each serving a unique purpose in the ecosystem. *Model repositories* store pre-trained serialized machine learning models and their metadata about their architecture and training data, making them readily accessible for downstreams tasks such as text generation, image classification, *etc*. *Dataset repositories* provide structured collections of data that can be used to train and evaluate these models, with easy access through Hugging Face's APIs. *Spaces* allow users to build and share interactive web applications, enabling developers to create demos and tools that others can easily explore and interact with, all hosted directly on the Hub.

Since we are interested in studying the evolution of model serialization formats used on Hugging Face, we focus on retrieving the serialized models that are available on *model repositories*. Thus, to collect the data needed for our study, we first used the Hugging Face Hub API [26] to retrieve the metadata for all model repositories' available on Hugging Face. This metadata includes information such as the repository's URL, its tags, a list of current filepaths, author, creation date, modification date, *etc*. At the time of our query (August 2024), the API returned a list of **996,939** model repositories. Henceforth, for the sake of simplicity, we will use the term "repositories" and "model repositories" interchangeably.

### C. Model Repository Filtering

We filtered this list to identify model repositories that match all of the following conditions[1]: **(i)** it was created before September 2022, **(ii)** last modified in 2024, and **(iii)** it contains at least one serialized model file. We narrow our search in this manner because safetensors was initially released in September 2022, and we are focused on observing how these repositories changed in response to safetensors being released to examine when developers started to shift from other (insecure) serialization formats to a safe one. While this filtering resulted in 1,034 repositories, 2 of them were gated, *i.e.*, a restricted-access repository where users must request and be granted permission to access the model. As such, our study includes **1,032** model repositories.

[1]While Hugging Face has models dating back to 2019, the earliest creation date in the collected metadata is **2022-03-02**. This is because the API only started storing creation dates after this point. Thus, this placeholder date is assigned to all models created before creation dates were being stored [27].

### D. Commit Logs Extraction

We extract the commit logs for all of these 1,032 repositories, obtaining a total of **17,773** commits[2]. We filtered these commits to identify those that contain *serialized model files*. Since a markdown file (*e.g.*, *README.md*) do not contain serialized models, we filter these out. Similar to a prior study [16], we include only commits that that modified, changed or added files with the following extensions: *bin*, *h5*, *hdf5*, *ckpt*, *pkl*, *pickle*, *dill*, *pth*, *pt*, *model*, *pb*, *joblib*, *npy*, *npz*, *safetensors*, and *onnx*. This led to a total of **4,889** commits to further analyze.

For each commit, we extracted the following metadata: the *repository url*, the commit's *hash*, *author*, *date*, *message*, as well as a *list of changed files*, and a list of *all of the files in the tree*. This allows us to see when certain files were added to a repository, and helps us to later identify when a particular model format was added to the repository.

### E. Serialization Method Identification

After extracting these commit logs, we clone each repository at each commit hash to identify the *serialization method* used for each model file in the repository. To do so, we ran a git checkout for the commit hash and extract all the model files in the repository. Subsequently, similar to a prior work [16], we employ a rule-based algorithm to identify the serialization method. The following are the rules (**R1**–**R5**) applied, in order:

**R1** **NumPy**: if the file's first 6 bytes is equals to the magic string `\x93NUMPY` [28].

**R2** **H5/HDF5**: if the file's first 4 bytes are equal to the string `\x89HDF` [29].

**R3** **safetensors**: if the first 8 bytes are equals to a 64 bit unsigned integer (*n*) and the remaining *n* bytes are a well-formed JSON [13].

**R4** When the first four bytes are equals to `PK\x03\x04`, then this is a zipped file that may contain one or more files that were saved using PyTorch or NumPy. We distinguish between these three formats by performing the following:
- **torch.jit.save**: if the zip file either contains a `constants.pkl` file or it contains a `code` folder with a `__torch__` subfolder.
- **NumPy**: if one of the files starts with the magic string `\x93NUMPY`.
- **torch.save**: if none of the above occurred, then the file was saved using PyTorch's method.

**R5** If the file starts with the hexadecimal value `\x80`, then the file was created using Pickle or a Pickle-based format, such as Dill, JobLib, or PyTorch.
- **Dill**: if the file contains the string `dill._dill\x94\x8c`
- **JobLib**: if the string *joblib.* is present in the file
- **PyTorch**: if the string *'little_endian'* and *'protocol_version'* is present in the file.
- **Pickle**: if none of the above occurred, then the file was saved using Pickle.

Once we determine the format of the model file, we save the repository URL, the commit hash, the model file path, and the determined serialization format to a CSV.

### F. Retrieving PRs created by Hugging Face's conversion tool

We extracted PRs created by Hugging Face's conversion tool from two sources:
- **Conversions dataset** [30]: Prior to April 2023, conversion usages were tracked and saved into a public dataset [30]. Thus, we download this dataset to obtain a list PRs and their associated metadata, such as status (open, closed, *etc*) and discussions. This dataset contained a total of **8,333** PRs.
- **SFConvertBot's community activity [31]**: Starting on March 2023, Hugging Face's conversion tool was modified to open PRs using a dedicated account (SFConvertBot [32]), instead of opening a PR using the account of the user that made the conversion request. As such, we wrote a web crawler that visit the SFConvertBot's community activity [31] and collect a list of pull request URLs. This crawler extracted a total of 43,596 PR urls. For each of these PR links, we retrieve their web pages to extract their metadata, namely developers' discussion (if any), the PR's merge status (*i.e.*, open, closed, merged), and any merge conflicts. This resulted in a total of **43,248** PRs.

### G. Merging & Filtering tool's PRs

After retrieving the PR from these two sources, we first merge the duplicated PRs. Next, we excluded PRs for which we could not obtain metadata (*i.e.*, cases in which our crawler received an HTTP response with a `4XX` status code). Moreover, we excluded the SFConvertBot's internal activity, *i.e.*, activity within Hugging Face's own repositories that do not correspond to a conversion request. This filtering step resulted in a total of **43,596** included in our study.

### H. Searching Developer Forums

To analyze developers' sentiments towards safetensors, we examined questions posted on StackOverflow, a popular platform where developers have discussions and ask questions regarding a variety of topics. To find all discussions relating to safetensors, we used the Stack Overflow API [33] to issue the following query: `"safetensor"` **`OR`** `"safe tensor"` **`OR`** `"safetensors"`. This search resulted in **268** posts. Similarly, we used the GitHub API [34] to find all pull requests that *(i)* contain the same aforementioned keywords, *(ii)* have been closed and *(iii)* were not opened by a GitHub bot. This search resulted in a total of **4,174** pull requests. Similarly, we also extracted discussions posted on the conversion tool's discussion board [35]. As such, we obtained a total of **24** discussions on Hugging Face.

### I. Sampling StackOverflow Posts and GitHub PRs

While discussions posted on Hugging Face's converter discussion board are guaranteed to be around safetensors and its usages, the posts and PRs collected from StackOverflow and GitHub, respectively, may contain discussions unrelated to safetensors. This is because the StackOverflow posts and GitHub PRs were obtained by a simple keyword-based search, the results may include false positives. Thus, we vectorized the text within StackOverflow posts and GitHub PRs using TF-IDF (Term Frequency–Inverse Document Frequency) [36] and computed the cosine similarity between the vectorized text and the query *"model serialization safetensors"*. Then, we sample posts and pull requests using a **95%** confidence level and a **5%** margin of error. Thus, we sort the posts/PRs by cosine similarity (descending) and sample the top 159 StackOverflow posts, and the top 352 pull requests from GitHub.

[2]We run this extraction on August 31st, 2024.

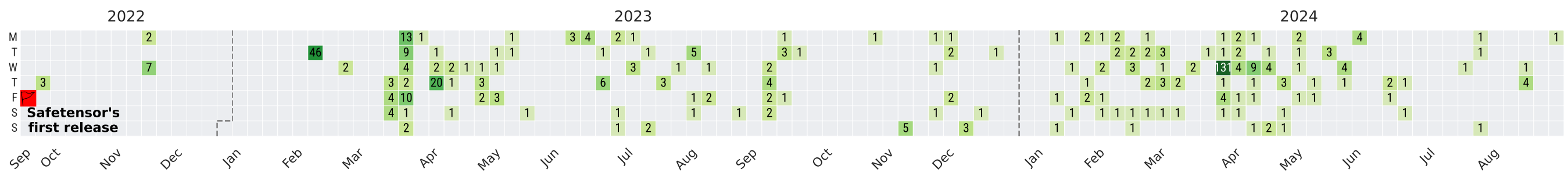

Fig. 4. RQ1: Calendar view of commits adding model serialized using safetensors on Hugging Face since September 2022

After this sampling, we manually analyzed the sampled GitHub PR and StackOverflow posts to identify true positives, *i.e.*, the post/PR is relating to or mentions safetensors, whether it is in the comment of the post, the title, or the body. During this step, we excluded GitHub PRs from safetensors' repository[3] where its code is hosted. We excluded these PRs because while the 'safetensors' term would come up a lot in these pull requests, they would not contain *developer perceptions*, but rather information related to the safetensors development itself. In the end, we obtained a total of **418** posts/discussions/PRs to analyze.

### J. Open Coding

We performed open coding on all the 418 collected samples. In this open coding process, we analyzed each post/pull request/discussion and annotated them with concepts, or codes [37], [38]. This open coding was performed by two authors of the paper, both with 1-2 years of software engineering experience. To asses the degree of inter rater reliability, we calculate the Cohen's Kappa [39] and have an agreement of **0.88**. This is a near perfect agreement.

Once this information was collected, we identified the key concepts from each post, and went through the process of refining these concepts/categories, resulting in broader categories which grouped these concepts into themes. The outcome is a ***taxonomy of developers' views on safetensors***. This taxonomy allows us to answer our RQ4. By finding the general patterns among these posts, we can identify developers' perspectives and thoughts regarding safetensors.

## IV. Results

This section presents the key findings from our analysis.

[3]https://github.com/huggingface/safetensors

### A. RQ1: Safetensors Usage Timeline

To answer this question, we inspected the commits that were *adding* safetensor files into existing model repositories. During this examination we found that only **418** (**9%**) commits out of **4,889** commits were adding models serialized using safetensors. Figure 4 shows a calendar heatmap in which each square represents a day and the numbers in them indicate how many commits added models serialized using safetensors. As shown in this figure, there were **3** commits that adopted the safetensors format only **6 days** after its first release on September 23rd, 2022. These commits were made by the same author in two model repositories (*openai-community/gpt2* and *FacebookAI/roberta-base*). This commit author is not only a member of both organizations (*openai-community* and *FacebookAI*) but also contributed to the safetensors project on GitHub[4]. This explains the quick adoption of safetensors shortly after its release.

We also found that out of all **418** commits adding safetensor files, most of them (**95.7%**) were commits to merge the pull requests made by Hugging Face's conversion tool [32]. That is, developers have used this tool to automatically convert their existing serialized PyTorch model file to the safetensors format. Figure 5 shows the distribution over time of commits merging PRs made by Hugging Face's conversion tool versus commits that were not merging the tool's PRs[5]. As shown in this figure, most of the non-merging commits are made *after* November 2023 (1 year and 2 months days after safetensors' first release). In between September 2022 through November 2023, developers mostly made use of Hugging Face's automated conversion mechanism to migrate their legacy PyTorch models to safetensors, which is a safer and more efficient serialization option [14].

We examined the data to verify whether developers were deleting the old PyTorch model files after using the conversion tool. Upon closer inspection of the commit logs, we noticed that **394** repositories (**98.7%**) still kept their old PyTorch model files in their repository.

Figure 5 also shows that there was a peak on March 27th, 2024. On this date, the Hugging Face's conversion tool has made **131** conversions. Although we were unable to find any concrete evidence for why the spike occurred, we hypothesize that this is due to Hugging Face releasing a newer version of the tool that addresses security issues found by HiddenLayer researchers. On February 21st, 2024, HiddenLayer researchers published an article [40] on how one can hijack the safetensors conversion space on Hugging Face by taking advantage of the way the model is loaded (using *torch.load*) when the model is being converted.

[4]https://github.com/huggingface/safetensors/graphs/contributors

[5]Given that the data has outlier data points (*i.e.*, dates with way more commits than the average), the y-axis uses a logarithmic scale to make it easier to see the markers.

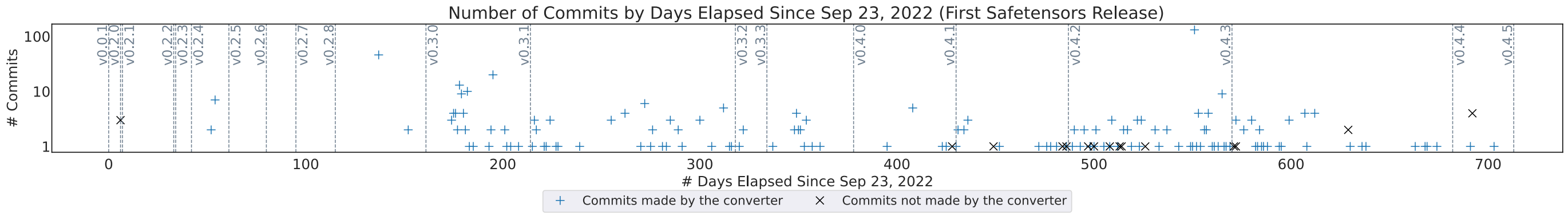

Fig. 5. RQ1: Distribution of commits made using Hugging Face's conversion tool

**RQ1 Findings Summary:**

- Model repositories have started adopting safetensors **6** days after its first release.
- Most of safetensor adoptions (**95.7%**) up to date were made via automated conversions of existing serialized PyTorch models using Hugging Face's conversion tool.
- Most repositories (**98.7%**) retained their old PyTorch model files even after using Hugging Face's conversion tool to convert these files to the safetensors format.

### B. RQ2: Safetensors' Adoption Evolution

To better examine safetensors' adoption trends, we first identify the *last* commit per repository per year. Subsequently, we group these commits by year then count yearly occurrences of different serialization formats. Figure 6 shows the distribution of different serialization formats over the years studied in this paper. We found that safetensors started gaining more adoptions in 2023.

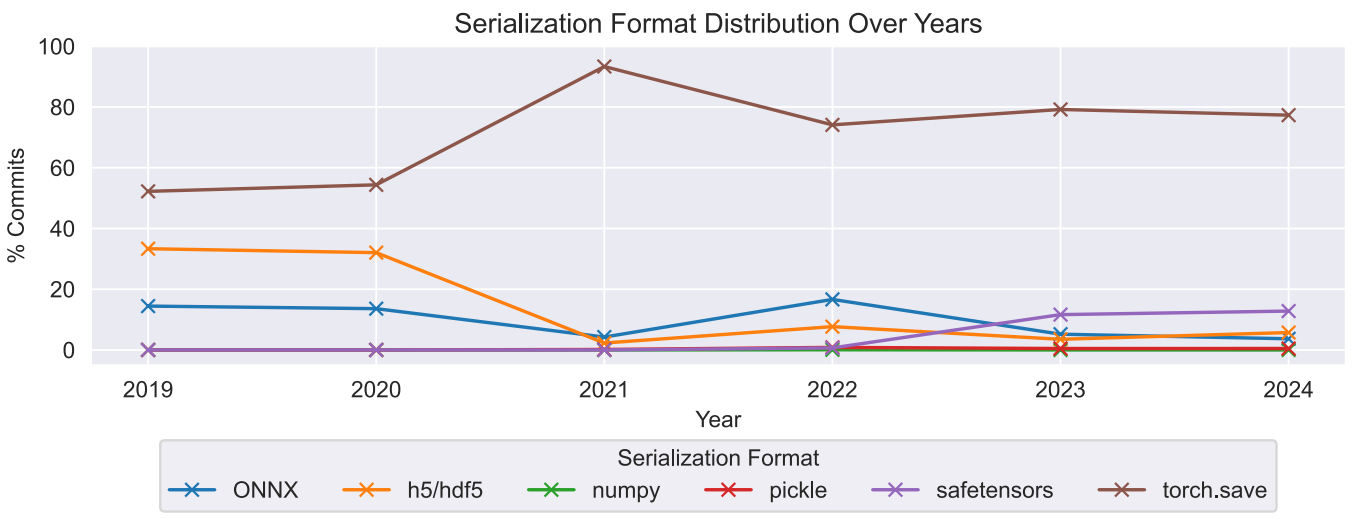

Fig. 6. Trend of serialization methods used over the years

As safetensors grows in popularity, it is replacing *torch.save* as a method of serialization. We observe that the number of *torch.save* files decrease, but the percentage of other serialization formats remain more or less constant. This suggests that developers are continuing to move towards using safetensors over other insecure methods of serialization, such as the pickle-based format offered by PyTorch.

These trends show that developers are willing and ready to adopt new, secure methods of serialization, although there is an initial hesitation towards using a new method. Additionally, it demonstrates that developers are aware, or at least willing to address security issues when it comes to serialization formats.

Interestingly, we observe that *torch.save* is the most popular serialization format across all years (2019-2024), with the h5/hdf5 format decreasing in popularity, and ONNX also decreasing in popularity, with a slight resurgence in 2022 before decreasing again. H5/HDF5 model formats, associated with tensorflow models, has been steadily losing popularity overtime. Since its release, Tensorflow was viewed as a difficult library to work with, with the learning curve on how to use it being as difficult or steeper than PyTorch or other ML frameworks [16], [41]. Thus, the observed decline in popularity of the h5/hdf5 model format is expected as tensorflow becomes obsolete.

ONNX had a brief boost in popularity in 2022, likely due to improved features being released [42], such as improved support with NVIDIA and Intel hardware, improved documentation, and general improved operability between frameworks. Nonetheless, *torch.save* remains as the dominant and preferred method of serialization to this day. The dependence on *torch.save* as a method of serialization still leaves a majority of model files on Hugging Face vulnerable to model injection.

**RQ2 Findings Summary:**

- Safetensors started to gain traction in 2023, and continues to gain popularity in 2024.
- Although safetensors is gaining popularity, *torch.save* (*i.e.*, PyTorch's insecure method of serialization) is still the primary method of serialization used, while other methods (excluding safetensors) are losing popularity.

### C. RQ3: Hugging Face's Conversion Tool Effectiveness

We found that developers used Hugging Face's Conversion tool to convert PyTorch models across **35,094** repositories. The first recorded conversion activity occurred on November 7th, 2022, which is approximately 1 and a half months after safetensors' first release. Figure 7 shows the distribution of the PRs' statuses on Hugging Face. Across all years, a majority of pull requests remain open (**83.5%**), while only **13.9%** of PRs are actually merged into the repositories. In 2024, however, more of the pull requests have been merged, with an increase from **12.7%** in 2023 to **30.6%**. An important distinction is that pull requests which were made by Hugging Face staff have a very high percentage of merged PRs (**93%**), while developers who are not a part of the Hugging Face staff tend to leave pull requests open (**84.3%** of PRs). This demonstrates that internally, Hugging Face employees trust and use this tool regularly, while the general population of developers either do not trust or do not want to use the conversion tool.

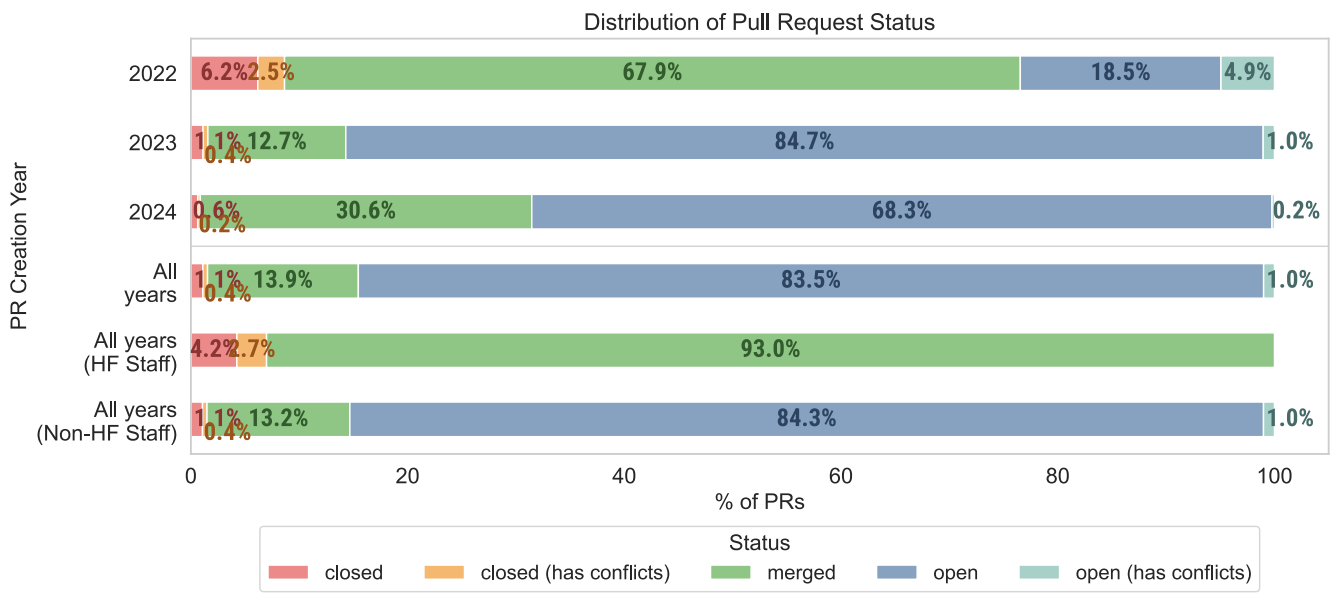

Fig. 7. RQ3: Distribution of PR Status Across Different Repositories

To better understand the reasons as to why developers were not merging these pull requests, we retrieved any PR that contained two or more developers' comments. We found **27** PRs that contained at least two comments from human developers (*e.g.*, not including the comment from the SFConvertBot). In these PRs, we found that the verification of the safetensors model file was a key concern (*i.e.*, testing the converted file to confirm it was not broken or behaving incorrectly). This indicates that developers are cautious before merging the PR. Other PRs suggest that this check is necessary, as they cite issues with the conversion, such as the model crashing on load due to a corrupted file, the conversion not creating the correct files, decreased performance when using the safetensors variant, or the file failing to load. Additionally, one discussion touched on previous issues with the safetensors conversion, citing that the conversion tool exported the file, but every time it was used it resulted in output from the 'normal' model, not the refined model.

One PR, in a repository run by Hugging Face developers, discussed automating the 'double-checking' that developers do to verify that the conversion to safetensors did not affect the original model's results. This double check is noted to be important, as safetensors' tied weights are managed differently from PyTorch. The developers confirmed that the tool does double-check inference code, but is not checking hidden states.

Some PRs' discussions were around how the safetensors file was created, and what safetensors is and why one should use it. One developer in particular criticized the conversion tool, along with questioning the point of safetensors in general. Hugging Face developers responded, explaining what safetensors is, and describing that the developer should understand *"the seriousness of what pickle files enable an attacker to do on your machine. And what an attacker could do if he got hold of anyone with write credentials in your company by modifying those weight files to everyone that is using your models."* Once again, this response highlights the importance of secure serialization formats, as well as the need to protect models from adversaries.

Interestingly, in one response to the question of why safetensors should be used, a developer shares that *"... the point [of using safetensors] is not so much in the "precaution" [from vulnerabilities]. Safetensors are loaded much faster into WebUI [...]"*. This indicates that this developer is not concerned with the safety of the file, but rather the speed with which it loads and can be deployed in their system. While the Hugging Face developers themselves recognize the severity of the issue regarding *security*, outside developers do not seem to be as concerned with the threat that pickle-based files pose.

**RQ3 Findings Summary:**

- Most pull requests (**83.5%**) made by Hugging Face's conversion tool remain unmerged.
- Some developers cited issues with the results of the conversion tool, with either the model not working as expected, or failing to load.

### D. RQ4: Developers' Perception

From our open coding of developers' discussions on StackOverflow, GitHub and Hugging Face, we identified **five** main discussion themes. Figure 8 compiles these finds as a taxonomy of specific concepts organized per discussion theme.

*Discussions on (Safe) Serialization:* Developer discussions reveal a strong interest in *safe serialization* practices. In particular, safetensors is contrasted with pickle formats, showing that developers are aware and concerned about the threat posed by pickle serialization methods. However, one pull request discussed *disabling safe serialization*, as the model could not support safetensor serialization. This demonstrates the need for developers to update their repositories to avoid the vulnerabilities associated with insecure methods of serialization.

Another pull request comes from the safejax repository [43], and focuses on *flax serialization*, specifically, how to serialize flax model parameters with safetensors. In other words, this pull request focuses on serializing flax models safely, using safetensors. Flax is a neural network library and ecosystem for JAX, designed by Google. Flax is currently used in the Alphabet research community [44].

In one post related to *model formats*, one developer highlighted the security concerns relating to pickle, and suggests that industry is likely to follow the shift towards the safetensors format in an effort to circumvent these concerns.

*Updating Safetensors:* Discussions revealed an ongoing effort to enhance safetensors' utility through adding more functionality and usability across a variety of repositories (*e.g.*, the safejax repository mentioned above, incorporating *flax serialization* using safetensors). *Safetensors pretrained* involves updates made to repositories which use or develop pretrained models. These updates are to enhance the usability of safetensors for pre-trained models. One pull request focused on adding support for *compressed safetensors*, which, as the name suggests, is a compressed version of the original safetensors format, developed to reduce the size of safetensor model files [45]. Some PRs discussed *refactoring safetensors*, *i.e.*, updating the use of safetensors (or the tensors themselves) within a repository.

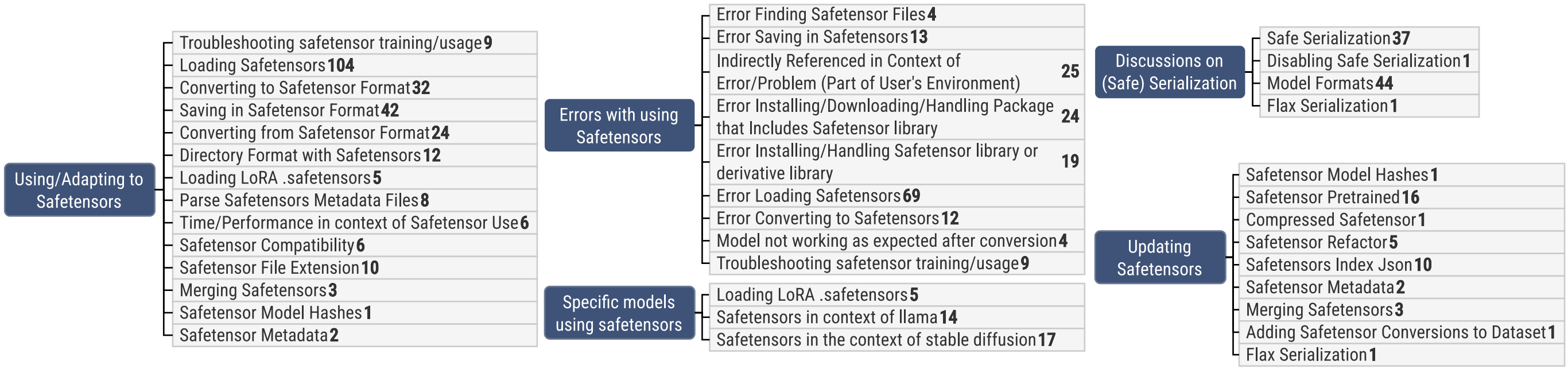

Fig. 8. Taxonomy of Developers' Perceptions

Additionally, developers are exploring ways to enhance the organization and accessibility of their models by creating or modifying *safetensors' index JSON* files. These indexed JSON files act as guides to the structure and content of a safetensors model, allowing users to quickly locate and retrieve specific model components.

Finally, one discussion from Hugging Face discussed *adding the safetensor conversion data to a dataset on Hugging Face*. This indicates Hugging Face's intention to understand how many developers and repositories are converting to safetensors.

*Specific models using safetensors:* Some developers discuss using specific models which are saved in safetensor files, such as LORA (Low-Rank Adaptation of Large Language Models; a lightweight training technique that reduces the number of trainable parameters [46]) and stable diffusion. These specific use-case questions, specifically the frequency of them, indicate that popular models are using safetensors as their method of serialization.

*Using/Adapting to Safetensors:* Developer discussions around safetensors reveal a concerted effort to understand and integrate this format into practical applications. One prominent focus is on how to *load safetensors*, highlighting initial difficulties with and solutions to adapting to the format. Discussions around *troubleshooting safetensor usage*, demonstrate the commitment to address practical challenges and improve usability in real-world scenarios. Additionally, developers are discussing the *time/performance of using safetensors*, which indicate that developers are aware of safetensors' efficiency and are looking for ways to enhance the speed and performance of their own applications.

Other concepts, such as *converting to/from safetensors format*, *safetensor compatibility*, *saving in safetensors format*, and *safetensor file extension* underscore the practical considerations that developers face when migrating to a new format like safetensors. Additionally, there were several posts involving *safetensor metadata*, where developers are learning how metadata is handled in the safetensors format, so that they are able to migrate their systems to safetensors more efficiently and effectively.

Some posts and pull requests surrounded concerns regarding *directory formats* and model selection protocols; in particular, prioritizing safetensor files when other model files are present in directories. This demonstrates a push towards standardizing the use of safetensors in model development. One pull request in particular discussed pre-calculating *model hashes* of safetensors files after training in an effort to speed up loading time for models and improve the efficiency of safetensor use. While this PR is not in the safetensors repository, it shows that developers are also looking to improve safetensor efficiency for their own use. Finally, some pull requests surrounded *merging safetensors*, in particular fixing issues that arose when attempting to merge safetensor model files to one.

*Errors with using Safetensors:* A number of developer discussions highlight various errors encountered when working with safetensors, reflecting both technical challenges and growing pains associated with adopting a new format. Some errors involved environment issues, which are categorized under *errors installing or handling the safetensors library or derivative library*. These issues typically involve common hurdles associated with installing new libraries.

Other issues involve using safetensors models, such as *errors finding the safetensors file*, *errors saving the safetensors file*, *errors loading the safetensors file*, *troubleshooting issues*, and *errors converting a file to safetensors*. While these issues could indicate regular growing pains with regards to a new format, a more concerning error is the *model not working as expected after being converted to safetensors*. The way a model is saved should never impact the model performance, and this error suggests a concerning issue that safetensors may actually be impacting the integrity of model weights. This highlights an area for further investigation, as any unintended alterations to model performance after serialization could limit the format's reliability and usability in production environments.

**RQ4 Findings Summary:**

- Many discussions surround how to load safetensors, as developers adapt to the new format
- While certain errors from using safetensors indicate adjustments to the new format, others indicate a deeper issue within the format (*e.g.*, models not working as expected)

## V. Discussion

This section provides the implications of our results to both software engineering researchers and practitioners.

### A. Developers' Pain Points

Our RQ4 results shed a light on developers' pain points when it comes to adopting safetensors. We observed that developers faced numerous errors relating to the basic use of safetensors, such as loading and saving safetensor files. These issues demonstrate the difficulties with integrating new formats into a development pipeline. A larger issue posed by safetensors is the threat of *model reliability*. Developers reported having performance issues in their models after converting to safetensors. This suggests that, in some cases, safetensors' serialization process impacts model integrity. This poses a big threat, as the serialization process should not alter model behavior. If safetensors is to be used reliably in practice, developers should have the certainty that their models will not be impacted by choosing a secure serialization format.

Additionally, although Hugging Face's tool (§ II-D) has facilitated the conversion of PyTorch models to safetensors, a majority of the pull requests created by the tool remain unmerged. Such an observation indicates possible hesitation or a lack of trust among developers for either the format or the tool. Many developers still manually verify that the tool's conversion did not alter their model, suggesting usability gaps in this automated tool.

### B. Implications for Researchers

Our findings have several implications for research, particularly as it pertains to the adoption and usage of safetensors. These implications serve as key directions for future works.

*Model Fidelity and Serialization:* Serialization should, in principle, preserve model fidelity, ensuring that the model's behavior does not change simply because it was saved and loaded. However, our findings in RQ3 and RQ4 demonstrate that developers are encountering unexpected performance issues in their models when they convert their existing models to safetensors. Not only does this impact developer's models in production, but it also impacts developer's ability to trust the tools given to them to use secure serialization. If serialization introduces performance degradation, it poses a significant barrier to the adoption of safe formats, like safetensors. Therefore, additional future works are needed on improving the reliability of automated convertion tools. Moreover, future research should investigate the potential impact serialization has on model integrity, and how to mitigate these issues. This includes investigating how data is handled during serialization, as well as compatibility across model architectures.

*Security and Usability in Serialization:* Our findings demonstrate that developers are increasingly motivated to adopt secure serialization formats in an effort to mitigate the security risks that are associated with pickle-based methods of serialization. However, a key point of safetensors that may be attracting more developers is its *performance enhancements* over other serialization formats. This dual appeal of security and efficiency suggests that, although security remains a key focus for researchers in developing serialization formats, usability and operational efficiency are also crucial for broad adoption. To encourage the widespread use of any secure serialization format, future research should address both security flaws and more practical benefits that developers can realize in production environments, making secure serialization a more practical choice for developers in real-world scenarios.

*Need for Enhanced Documentation and Support:* Given the vast number of issues developer face while trying to adopt safetensors, improved documentation and dedicated support channels for learning and troubleshooting issues with safetensors could assist developers in making the adjustment. Developers would benefit from better resources on handling common issues, such as the basic usage of safetensors (*e.g.*, loading and saving models), format conversion, model verification after conversion, and environment compatibility. Thus, future research is needed on automated tools and plugins that integrate with existing developer environments to simplify and guide the adoption of safetensors. Ensuring that developers have the documentation they need to integrate a secure serialization format into their pipelines allows for these methods to be more readily and easily adopted.

### C. Implications for Practitioners

*Don't trust all models on Hugging Face:* A majority of models on Hugging Face still use unsafe serialization methods. Therefore, it is important that developers do not blindly trust these models, as they can potentially be malicious, and Hugging Face's security scanner may not accurately alert developers of the threat or any hidden payloads [16]. Before developers download and load models, they should verify that the authors are known and trusted, and load the model in a secure environment (*e.g.*, Docker [47]) to prevent any unexpected exploits from harming their systems.

*Delete Old Files:* As practitioners migrate to safetensors, it is important to delete old serialized files that use insecure methods of serialization to protect users from the threats that come with using them. Keeping the older serialized files in production environments still allows an attacker to take advantage of them, and puts downstream users at risk of accidentally loading a malicious model. Moreover, redundant files can lead to confusion over which model version is up-to-date. For example, maybe the `pytorch_model.bin` file was updated with new weights, but the `model.safetensors` file remains unchanged. Users would get different results from both of these files, thus increasing the risk of using outdated models, or unsafe models unintentionally. By removing old files, practitioners can reduce the amount of storage used, mitigate security risks, and ensure that only trusted, updated files are deployed.

*Leverage Automation but Verify Outputs:* Our RQ3 results show that developers are not merging the pull requests created by Hugging Face's conversion tool. While automated tools are useful, practitioners should view them as a starting point, and should manually verify models before they are deployed in production so that their models do not behave unexpectedly. This approach balances the efficiency and convenience of the tool with the reliability of manual verification.

## VI. Related Works

### A. Object Injection

The security risks posed by built-in unsafe deserialization mechanisms provided by multiple languages (*e.g.*, have been widely studied [48]–[57]. Slaviero [12] was the first to publicly document the shellcoding techniques that are possible when targeting Python's pickle module, highlighting the need for robust defenses against Pickle-based attacks. Huang *et al.* [11] explored how developers implement a common defense mechanism called *"restricted unpickler"* and developed an approach to detect and exploit flawed implementations of this defense mechanism. David *et al.* [58] developed QUACK, which mitigates deserialization attacks in PHP applications by using static analysis to identify safe classes.

### B. Empirical Studies on Software Evolution

There have been a number of studies investigating the ways in which software ecosystems evolve. In this regard, Bavota *et al.* [59], [60] investigated dependency evolutions within Apache focused on identifying the motivation behind dependency upgrades, and specific factors which impact adoption patterns. Another study from Ringlstetter *et al.* [61] investigated the use of evolution annotations in Java open-source projects that utilize NoSQL mappers. They found that only a minority of projects employ these annotations for actual data evolution while a majority of developers often re-purpose annotations for tasks like serialization or time-stamping, indicating possible gaps in the tool's adoption for its intended purpose. Jaime *et al.* [62] introduced two novel metrics, "rhythm" and "speed", to examine release dynamics within the Maven ecosystem, highlighting an accelerating trend in release cycles that reflects a shift towards more frequent updates, particularly in response to critical events like vulnerabilities. D'Ambros *et al.* [63] used the Release History Database (RHDB) to study architectural decay, developer effort distribution, and change coupling within software systems. While most of these works focus on the evolutions or ecosystems of Java projects, our work investigates the evolution of model serialization formats for ML models developed in Python. We specifically investigate the evolution of models available on Hugging Face.

Tinnes et al. [64] proposed the RaMc approach, using retrieval-augmented generation and large model repositories to enable context-sensitive software model completions with high accuracy. While this work applies LLMs to support structure-driven model evolution, our study examines the adoption of safetensors, a secure serialization format, in ML systems, emphasizing security in model storage evolution.

### C. Model Security

Golla [65] surveys security and privacy vulnerabilities in deep learning models, categorizing threats as model extraction, inversion, adversarial, and poisoning attacks. The study highlights various attack methods and evolving tactics, such as advanced query strategies [66], [67] and stealthy backdoors [68], [69]. Taraghi *et al.* [70] explore the reuse of pre-trained models within Hugging Face, underscoring security risks as a critical challenge due to undocumented or unverified model behaviors that can expose developers to vulnerabilities. The study points to community-driven efforts to enhance documentation and metadata management as partial solutions to these risks, supporting safer reuse practices. Ho *et al.* [71] examine bug patterns in PyTorch, offering insights into common defects that affect deep learning libraries and their impact on developer workflows. While these studies provide valuable perspectives on model reuse, attack methods in machine learning, and bug resolution, our work uniquely focuses on the adoption of safetensors, emphasizing model format evolution rather than model usage or defect management.

## VII. Threats to Validity

We discuss our study's threats to validity around construct, internal, and external validity threats [72].

***Construct validity***: A threat to our study pertains to the interpretations of developers' sentiments, as well as the categorization and coding of posts. Thus, our analyses rely on the accuracy and consistency of the reviewers' categorizations during open coding, which inherently involves some level of subjectivity. To mitigate potential biases in categorization, we conducted independent reviews of the data.

***Internal validity***: Due to the potential for incorrect categorization of posts, we had the two authors performing the categorization independently review each post and then resolve any discrepancies. To assess the reliability of our categorization, we calculate the Cohen's Kappa and have a score of **0.88**, indicating a near perfect agreement among our reviewers [73].

***External validity***: Our study is limited to Hugging Face, thus we may miss developers using safetensors on other platforms such as GitHub or Kaggle. However, given Hugging Face's position as one of the largest and most popular model-sharing hubs, we likely captured a majority of safetensors usage through our analysis.

## VIII. Conclusion

We examined developers' adoption, perceptions, and challenges surrounding safetensors through an analysis of community discussions, usage patterns, and automated conversion tools. Our findings reveal a growing interest among developers in secure serialization methods, with safetensors gaining traction due to its promise of security and improved performance. However, the results also reveal significant pain points that developers face when adopting this format, including frequent errors, compatibility issues, and concerns over model fidelity after converting from other formats. The study highlights several key implications for practitioners, emphasizing the need for careful verification of model performance after conversion to safetensors. For researchers, the findings underscore the importance of reliable tools and designing serialization formats that ensure both security and operational efficiency.